\documentstyle[emulateapj]{article}

\newcommand{\kps}{\,\rm{km~s}^{-1}}

\newcommand{\lsun}{\,L_{\odot}}

\begin{document}

\lefthead{Frayer et al.}

\righthead{Near-Infrared Colors of Submillimeter Galaxies}

\submitted{Accepted (22Oct03) and scheduled for publication in AJ (Feb04)}

\title{Near-Infrared Colors of Submillimeter-Selected Galaxies}

\author{D.\ T.\ Frayer\altaffilmark{1},
N.\ A.\ Reddy\altaffilmark{2},
L.\ Armus\altaffilmark{1}, 
A.\ W.\ Blain\altaffilmark{2},
N.\ Z.\ Scoville\altaffilmark{2},
Ian Smail\altaffilmark{3}}

\altaffiltext{1}{SIRTF Science Center, California Institute of Technology
220--06, Pasadena, CA  91125, USA} 
\altaffiltext{2}{Astronomy Department, California Institute of Technology
105--24, Pasadena, CA  91125, USA} 
\altaffiltext{3}{Institute for Computational Cosmology, University of Durham, South
Road, Durham, DH1 3LE, UK}

\begin{abstract}

We report on deep near-infrared (NIR) observations of
submillimeter-selected galaxies (SMGs) with the Near Infrared Camera
(NIRC) on the Keck~I telescope.  We have identified $K$-band candidate
counterparts for 12 out of 15 sources in the SCUBA Cluster Lens Survey.
Three SMGs remain non-detections with $K$-band limits of $K> 23$\,mag,
corrected for lensing.  Compensating for lensing we find a median
magnitude of $K=22\pm1$\,mag for the SMG population, but the range of
NIR flux densities spans more than a factor of 400.  For SMGs with
confirmed counterparts based on accurate positions from radio, CO,
and/or millimeter continuum interferometric observations, the median NIR
color is $J-K=2.6\pm0.6$\,mag.  The NIR-bright SMGs ($K<19$\,mag) have
colors of $J-K \simeq 2$\,mag, while the faint SMGs tend to be extremely
red in the NIR ($J-K>3$\,mag).  We argue that a color selection
criterion of $J-K\ga3$\,mag can be used to help identify counterparts of
SMGs that are undetected at optical and radio wavelengths.  The number
density of sources with $J-K>3$\,mag is 5\,arcmin$^{-2}$ at
$K<22.5$\,mag, greater than that of SMGs with S(850$\mu$m)$>2$\,mJy.  It
is not clear if the excess represents less luminous infrared-bright
galaxies with S(850$\mu$m)$\la 2$\,mJy, or if the faint extremely red
NIR galaxies represent a different population of sources that could be
spatially related to the SMGs.

\end{abstract}


\section{Introduction}

The identification of the optical counterparts of the high-redshift
population of submillimeter galaxies (SMGs) (Smail, Ivison, \& Blain
1997; Hughes et al.\ 1998; Barger, Cowie, \& Sanders \ 1999a; Eales et
al.\ 1999, 2000; Cowie, Barger, \& Kneib 2002; Chapman et al. 2002a;
Scott et al. 2002; Smail et al. 2002; Blain et al. 2002; Webb et
al. 2003) is challenging because of the large beam size of the 850$\mu$m
detections and the general faintness of their rest-frame ultra-violet
continuum emission (e.g., Smail et al. 2002).  The most successful
technique has been to use accurate radio positions to identify candidate
counterparts (Smail et al. 2000; Barger, Cowie, \& Richards 2000; Ivison
et al. 2002; Chapman et al. 2001, 2003a).  Unfortunately, current radio
observations only detect about 50--70\% of the population (Smail et
al. 2000; Chapman et al. 2003a), of which most are at $z\sim2$--3
(Chapman et al. 2003b).  For sources that lack radio detections, we are
forced to alternative techniques for the identification of the
counterparts.  For example, Smail et al. (1999) identified two candidate
counterparts based on being extremely red objects (EROs, $R-K>6$\,mag).
EROs have been shown to contribute statistically to the 850$\mu$m
background (Wehner, Barger, \& Kneib 2002), suggesting a natural
connection between dusty SMGs and galaxies with extremely red colors.
However, the identification of ERO counterparts based on their $R-K$
colors is limited to only the brightest examples since most SMGs have
$K>21$\,mag and typical $R$-band limits are $R< 27$\,mag.  A similar
strategy can be applied using near-infrared (NIR) colors.  Recent deep
NIR surveys have shown the usefulness of the $J-K$ colors for the
identification of extremely red galaxies (e.g., Totani et al. 2001b,
Franx et al. 2003).  By using $J-K$ colors we can identify extremely red
galaxies in the SMG fields at fainter depths than currently achievable
by $R-K$.

In this paper we discuss deep NIR observations of SMGs in the SCUBA
Cluster Lens Survey (Smail et al. 1997, 2002).  This survey represents
sensitive sub-mm mapping of seven gravitational-lensing clusters that
uncovered 15 background SMGs.  The advantage of this sample is that the
amplification of the background SMGs by the clusters allows for deeper
source-frame observations.  Previous $K$-band observations of the fields
only reached depths of $K\sim21$\,mag (Smail et al. 2002), which was not
sufficient to identify about half of the SMGs.  The goals of these NIR
observations are straight-forward.  In fields with no clear $K$-band
counterparts, we observed much deeper in $K$-band to search for very
faint sources (e.g., Frayer et al. 2000).  In fields with candidate
$K$-band counterparts, we observed in both $J$- and $K$-band to measure
the $J-K$ colors of the population and to attempt to identify the most
likely counterparts based on extremely red $J-K$ colors.

A cosmology of H$_o=70\kps\,{\rm Mpc}^{-1}$, $\Omega_{\rm M}=0.3$, and
$\Omega_{\Lambda}=0.7$ is assumed throughout this paper.

\section{OBSERVATIONS AND DATA REDUCTION} 

We observed the SCUBA Lens Cluster SMGs using the Near Infrared Camera
(NIRC) on the Keck~I\footnote{The W.\ M.\ Keck Observatory is operated
as a scientific partnership among the California Institute of
Technology, the University of California, and NASA.  The Keck
Observatory was made possible by the generous financial support of the
W.\ M.\ Keck Foundation.}  telescope during seven nights between UT 1999
October and 2002 March.  The NIRC instrument uses a $256\times256$ InSb
detector with a pixel scale of $0\farcs15$ (Matthews \& Soifer 1994).
We observed using the standard $J$-band ($1.2\mu$m) and $K$-band
($2.2\mu$m) filters.  Exposures of 10\,s$\times$6\,coadds were taken in
$K$-band, while exposures of 20\,s$\times$3\,coadds were taken in
$J$-band.  To provide uniform coverage across the image, a random
dithered pattern was used with 10--15$\arcsec$ offsets between adjacent
exposures.  The seeing during the nights varied from about $0\farcs4$ to
$1\farcs0$ (FWHM).  Approximately 15\% of the data with relatively poor
seeing ($>0\farcs8$) or seriously affected by internal reflections of
Moon light were rejected.  Residuals due to the Moon partially affect
the sensitivity of the $J$-band data of SMM\,J00265+1710 and the
$K$-band data of SMM\,J00267+1709.  Table~1 summarizes the observing
runs and provides the effective integration time on each SMG field after
data editing.

We used the IRAF\footnote{IRAF is distributed by the National Optical
Astronomy Observatory, which is operated by the Association of
Universities for Research in Astronomy, Inc., under cooperative
agreement with the National Science Foundation.} NIRCtools package
(D. Thompson, private communication) to reduce the data.  Sets of dark
frames from each night were subtracted from each exposure to remove the
dark current as well as the bias level.  The dark-subtracted exposures
were divided by a normalized skyflat.  Frames were sky-subtracted using
temporally--adjacent images to yield reduced exposures.  The individual
exposures were aligned to the nearest pixel using common objects in the
frames.  The data were placed on the Vega-magnitude scale from
observations of a set of near-infrared standard stars (Persson et al.\
1998) taken at a range of airmasses to correct for extinction.  Based on
the dispersion of zero-points determined throughout the runs, the
uncertainty of the derived magnitude scale is about 0.05\,mag for large
apertures.

Data from multiple nights were combined with weights proportional to
their integration times.  For source identification and photometry, we
used the Sextractor program (Bertin \& Arnouts 1996).  The reduced $J$-
and $K$-band maps were combined to produce deeper NIR images for source
detection.  These combined images were used as the reference images to
define the aperture centers for photometric measurements on the
individual $J$- and $K$-band images.  Weight maps proportional to the
inverse of the local variance were used to yield representative errors.
The images were registered to the APM coordinate system using optical
data of the fields from Smail et al. (2002).  For the case of
SMM\,J04433+0210 (N4), we adopted the updated astrometric solution from
Neri et al. (2003).

\section{RESULTS}

Figure~1 shows the $J$- and $K$-band images of the SMG fields. The
extent of the images matches the FWHM beam of the SCUBA instrument at
850$\mu$m.  The images are centered on the sub-mm positions listed in
Table~1, which are accurate to about 3$\arcsec$ for the brightest sub-mm
sources.  For all but three sources, we adopted the sub-mm positions
given by Smail et al. (2002).  In the case of SMM\,J22471$-$0206, the
peak position of the sub-mm emission was used.  This is offset by about
4$\arcsec$ north-west from the position tabulated by Smail et al.\
(2002), and the uncertainty in the position of SMM\,J22471$-$0206 may
result from blending with a fainter sub-mm source.  For SMM\,J21536+1742
the position from the subsequent deeper observations from Cowie et
al. (2002) was used.  We also adopted an average central position for
SMM\,J02400-0134 based on the Smail et al. (2002) position and the
position of Cowie et al.'s source\,\#2.  Galaxies previously identified
are labeled with their name-ID given by Smail et al. (2002).  New
candidate SMG counterparts are designated with an ``A'' and additional
extremely red NIR ($J-K>3$\,mag) sources in the fields are designated
with a ``B''.

\subsection{Photometry Results}

Table~2 shows the results of aperture measurements for the confirmed and
candidate counterparts of the SMGs.  The confirmed SMG counterparts are
based on accurate positions from CO, millimeter continuum, and/or radio
interferometric observations (see references in Table~2).  The candidate
counterparts are discussed in \S3.2.  For the majority of the sources,
which are small, a 14-pixel-diameter aperture ($2\farcs05$) was used,
while larger apertures were used as appropriate for extended sources.
Aperture corrections, as a function of aperture size, were calculated
using the observations of standard stars.  Sources with blended
components were measured using Sextractor.  For the faint SMGs M12 and
N4, the bright nearby foreground galaxies were subtracted from the
images before aperture measurements were made (e.g., Frayer et
al. 2003).  For comparison with the known SMGs and to search for
possible new counterparts, all sources detected above the $3\sigma$
level were measured.  The images are about one to two magnitudes deeper
than the previous $K$-band observations of the fields (Smail et
al. 2002), and the photometric results in this paper are consistent with
those measured previously.

Figure~2 shows the $J-K$ colors as a function of $K$-band magnitude for
the 339 sources detected in the 12 fields with both $J$- and $K$-band
imaging.  The survey limits for the data are approximately $K<22.5$\,mag
and $J<24.5$\,mag ($3\sigma$).  Typical galaxies in the foreground clusters
($z=0.2$--0.4) are expected to have colors of $J-K\la 1.6$\,mag, while
the background galaxies at $z>1$ are expected to have colors of
$J-K>1.5$\,mag (e.g., Totani et al. 2001b).  The seven confirmed SMG
counterparts in the sample have $J-K$ colors ranging from 1.7 to
$>4.6$\,mag.  The bright SMGs ($K<19$\,mag) are bluer ($J-K\sim2$\,mag)
than the fainter SMGs which tend to show extremely red NIR colors
($J-K\ga3$\,mag).  The NIR results suggest that the $J-K$ colors can be
used to help identify candidate SMG counterparts.  Based on this sample,
bright, relatively blue NIR sources ($K<19$\,mag with $J-K< 1.5$\,mag),
and the fainter sources ($K>20$\,mag) out to a redder limit of
$J-K<2$\,mag are not likely to be the counterparts of typical SMGs.

Besides the confirmed and candidate SMG counterparts, we have uncovered
several additional sources in the fields which have very red
($J-K>2.5$\,mag) or extremely red ($J-K>3$\,mag) NIR colors.  All of the
extremely red sources are faint ($K>19$\,mag), presumably since bright
extremely red sources are rare (e.g., Totani et al. 2001b) and the
survey covers a small area.  In total, we find 13 sources with
$J-K>2.5$\,mag and 9 sources with $J-K>3$\,mag.  The reddest source in
the sample has $J-K>4.9$\,mag ($3\sigma$), which is the reddest galaxy
in the NIR known to date.  In comparison, the previous reddest galaxy
has $J-K>4.5$\,mag and was found in the Hubble Deep Field (Dickinson et
al. 2000).

\subsection{Individual Sub-mm Galaxies}

The identification of the counterparts of the SMGs in the SCUBA Cluster
Lens Survey has been discussed previously (Smail et al. 2002 and
references therein).  We now present details of the new candidate
counterparts and the additional knowledge gained from the deep NIR imaging of
the fields.

SMM\,J09429$+$4658 --- Smail et al. (1999) identified this source as H5
based on its extremely red $R-K$ color and an accurate radio position.
The deep $J$-band data for H5 indicate an extremely red NIR color of
$J-K>4.57$\,mag $(3\sigma)$.  This is the reddest known NIR counterpart
to a SMG.  The source B1, which is 4$\arcsec$ north-east of H5, also
shows an extremely red NIR color of $J-K>3.84$\,mag.  It is not known if
this source is associated with H5; neither source has a redshift.
However, the likelihood of randomly finding two $J-K\ga4$\,mag sources
within 5$\arcsec$ is less than 0.001 (Totani et al. 2001b), which
suggests the two galaxies could be related.  Unlike H5, B1 is detected
at optical wavelengths and is not an ERO based strictly on its $R-K$
color.

SMM\,J22471$-$0206 --- P1 has been considered a possible counterpart
based on its tadpole morphology (Smail et al.\ 1998).  The relatively
blue $J-K<2$\,mag color for P1 is circumstantial evidence against it
being the SMG counterpart.  Smail et al. (2002) and Barger et
al. (1999b) suggest that P4 may be the SMG counterpart based on its
morphology, red color, and the presence of an AGN.  The argument against
P4 is the fact that its high submm-to-radio flux density ratio is
inconsistent with its low redshift of $z=1.16$, assuming a standard dust
temperature of $T\sim40$--50\,K.  If P4 is the sub-mm counterpart, the
upper-limit on the 1.4-GHz radio emission implies a dust temperature of
less than about 22\,K (see Blain, Barnard, \& Chapman 2003), even cooler
than the two cold dusty galaxies found in the FIRBACK survey (Chapman et
al. 2002b).  No other new candidate sources were found in this field
from the NIR data.  P4 is currently the best candidate detected near the
sub-mm position given its relatively red color of $J-K=2.3$\,mag, but
its identification is not conclusive.

SMM\,J02400$-$0134 --- Smail et al. (2002) classified this as a blank
field, but deep NIR imaging has uncovered two possible counterparts.
The redder source A1 ($J-K=2.9$\,mag) is near the position of
source\,\#2 of Cowie et al. (2002), while A2 which is nearly as red in
the NIR ($J-K=2.5$\,mag) lies closer to the position of Smail et
al. (2002).  It is possible that the sub-mm emission arises from both
sources.  Given that the Cowie et al. (2002) SCUBA observations are
deeper, we suggest A1 is the most likely candidate SMG for the field.

SMM\,J21536$+$1742 --- The faint source K2 near the sub-mm position of
Smail et al. (2002) has a relatively blue $J-K=1.3$\,mag color and is
not believed to be the counterpart.  K3 shows a strong NIR color
gradient between components K3a ($J-K=2.3$\,mag) and K3b
($J-K=1.8$\,mag).  The sub-mm centroid from Cowie et al. (2002) is
positionally coincident with that of a weak ISOCAM source (object-ID 43,
Metcalfe et al. 2003) and is near K3a.  The red NIR color of K3a is
consistent with the identification of K3 as the SMG counterpart.

SMM\,J00265$+$1710 --- The source designated M13 (Fig. 1) is possibly
associated with radio emission (Smail et al. 2002), but its relatively
blue $J-K$ color suggests that it is not likely to be the SMG.  This SMG
is considered to be a non-detection down to $K>22.5$\,mag ($3\sigma$).

SMM\,J22472$-$0206 --- Although P2 ($K=21.9$\,mag) is at high-redshift
($z=2.1$, Barger et al. [1999b]), it is not thought to be the SMG based
on its blue $J-K<2$\,mag color (Table~2).  The reddest source in the
field P7 ($J-K>3.8$\,mag) is also nearest to the sub-mm position and is
adopted as the likely SMG candidate counterpart.  The nearby source A3
(Fig. 1) is very red as well ($J-K=2.7$\,mag) and may be associated with
P7.

SMM\,J04433$+$0210 --- This SMG is the weakest sub-mm source in the
survey and is possibly associated with a weak radio detection (Smail et
al. 2000) at the position of N5 (Smail et al. 2002).  We find a faint
$K$-band magnitude of $22.4$ for N5.  Since no other sources were
detected in $K$-band near the sub-mm position that were not already seen
in the $I$-band data, N5 remains the best candidate SMG counterpart.

\section{DISCUSSION}

The current status regarding the counterparts of the 15 SMGs in the
SCUBA Cluster Lens Survey can be summarized as follow.  Seven SMGs have
reliable counterparts (L1/L2, J1, M12, H5, J5, L3, and N4) based on
accurate positions from CO, radio, and/or millimeter interferometric
continuum observations (Frayer et al., 1998, 1999, 2000; Smail et
al. 2000; Ivison et al. 2000; J.-P. Kneib et al., private communication;
R. Neri et al. 2003).  An additional five SMGs have candidate
counterparts (\S3.2): P4 for SMM\,J22471$-$0206, A1 for
SMM\,J02400$-$0134, K3 for SMM\,J21536+1708, P7 for SMM\,J22472$-$0206,
and N5 for SMM\,J04433$+$0210.  The three remaining fields
(SMM\,J00265$+$1710, SMM\,J00266$+$1710, and SMM\,J0267$+$1709) are
still considered undetected to about $K \ga 23$ ($3\sigma$), after
correcting for lensing.  The faintest sub-mm sources in the survey may
be affected by confusion and/or may even represent spurious 850$\mu$m
detections in some cases.  The limitations due to confusion have been
previously discussed (e.g., Eales et al. 2000; Blain et al. 2002).
Seven of the nine brightest SMGs (Table~1) detected above $4\sigma$ at
850$\mu$m (Smail et al. 2002) have confirmed counterparts, and all nine
of the brightest SMGs have candidate NIR counterparts.

\subsection{Near-Infrared Properties of Submillimeter Galaxies}

Figure~2 shows that SMGs tend to be either relatively bright ($K\la
19$\,mag) with normal NIR colors consistent with high redshifts ($z>1)$
or faint with extremely-red colors ($J-K\ga 3$\,mag). The distinct color
difference between the bright SMGs and the faint extremely-red SMGs is
consistent with about 3~mag of additional source-frame visual extinction
for the fainter sources assuming typical SMG redshifts of $z\sim2$--3.
However, since the SMGs do not appear uniformly distributed along the
reddening vector (Fig.~2), reddening may not be the only explanation for
the variation of the NIR colors.  A dichotomy of colors for SMGs has
previously been seen in $I-K$ (e.g., Smail et al. 2002) and has been
used as a classification scheme where the Class-I SMGs are relatively
bright sources and the Class-II sources are faint and extremely red
(Ivison et al. 2000).  The NIR data are consistent with the presence of
two classes of SMGs based on colors.  There are many possible origins
which could contribute to the variations of colors including reddening,
the presence of an AGN, the presence of unobscured companions, and/or
different evolved stellar masses.

With several undetected SMGs, the average $K$-band magnitude is still
intermediate, but we can estimate the median $K$-band magnitude for the
SMG population.  We correct for lensing by using the amplification
factors given by Smail et al. (2002) and Cowie et al. (2002).  For
sources without known amplification factors, we adopt an average value
of 2.5.  Assuming all of the candidate SMG counterparts are correct, the
median $K$-band magnitude is $K=21.0$\,mag.  This result is likely
biased by the selection effect that it is easier to detect brighter
candidates and should probably be considered as a lower-bound.  An
upper-bound of $K<23.3$\,mag can be estimated by assuming all of the
candidate counterparts with $J-K<3$\,mag are actually still undetected.
The median magnitude of the SMG population is thus likely to be
$K=22\pm1$\,mag.

At the average redshift of about $z\sim 2.5$ for a significant fraction
of the SMG population (Chapman et al. 2003b), the observed $K$-band
emission represents rest-frame optical $R$-band light.  By comparing the
observed NIR properties of SMGs with the optical properties of
Ultraluminous Infrared Galaxies (ULIRGs, $L_{ir}>10^{12}L\sun$), we can
test the hypothesis that SMGs are analogous to the low-redshift
population of ULIRGs.  The median $K=22$\,mag for the SMG population
corresponds to an absolute magnitude of $M_{R}=-21.5$\,mag at $z=2.5$.
For the low-redshift {\it IRAS} 1-Jy ULIRG sample, the median $R$-band
absolute magnitude is $M_{R}=-21.9$\,mag, corresponding to $2L^{*}$
(Kim, Veilleux, \& Sanders 2002).  Therefore, SMGs have comparable
optical luminosities to local ULIRGs and thus likely represent massive
systems.

The median NIR color of the 6 confirmed SMG counterparts with meaningful
NIR colors is $J-K=2.6\pm0.6$\,mag.  Adding the additional candidate
counterparts does not change the median value.  However, given the
selection effect that brighter sources which tend to have bluer NIR
colors are easier to detect, the true median $J-K$ color of the SMG
population may be significantly redder.  We can compare the observed
$J-K$ colors with the corresponding rest-frame $U-R$ colors found for
ULIRGs.  We calculate a median color of $U-R=1.15$\,mag for the sample
of ULIRGs with global $U$-band measurements from Surace \& Sanders
(2000) and global $R$-band measurements from Kim et al. (2002).  At a
redshift of $z=2.5$, the median $U-R$ color of local ULIRGs would
correspond to a NIR color of $J-K=2.7$\,mag, in close agreement with the
SMGs.  Low-redshift ULIRGs show a broad range of colors, presumably due
to patchy extinction.  From the full range of $U-R$ colors found for
ULIRGs, we could expect colors of $J-K=1$ to 4.2\,mag for the SMGs.  We
have yet to identify any SMGs with colors as blue as the bluest ULIRGs,
but we have detected SMGs redder than the reddest known ULIRGs (e.g.,
H5).

In comparison, the Lyman-break population of galaxies (LBGs), which have
similar redshifts to the SMG population, are significantly bluer with an
average value of $J-K_{\rm s}=1.5$\,mag (Shapley et al. 2001).
Therefore in terms of rest-frame optical colors, SMGs are more similar
to low-redshift ULIRGs than typical LBGs.

Most SMG counterparts are not compact.  We fitted an elliptical Gaussian
to the $K$-band data to measure the FWHM of the core of the SMG
counterparts.  After deconvolution with the stellar
point-spread-function, the central FWHM of the SMG cores ranges in size
from $<0\farcs5$ to $2\farcs1$.  The median core size is $1\farcs3$ for
the seven confirmed SMGs, and is $1\farcs2$ including the candidate
counterparts.  Assuming typical SMG redshifts of $z=2$--3 (Smail et
al. 2002; Chapman et al. 2003b), the radius of the SMG cores is about
2\,kpc, corrected for lensing.  The sizes of the SMGs are similar to
those derived for the bright central regions of ULIRGs (Surace et
al. 1998; Scoville et al. 2000) and are comparable to the bulge sizes
found for normal galaxies.  These results are consistent with the
scenario that the SMGs represent the formative phases of the bulges of
galaxies or the cores of massive ellipticals (Lilly et al. 1999).

Several SMG counterparts show multiple components suggesting that
mergers play a significant role for the SMG population.  Nearly all
local ULIRGs have been shown to arise from mergers based on optical
morphologies that show strongly interacting pairs with apparent tidal
tails and distorted galaxies with close double nuclei (Murphy et al.\
1996; Surace et al. 1998).  The unusual optical morphologies are a
distinguishing characteristic of ULIRGs.  Unfortunately, the NIR data
for the SMGs lack both the surface brightness sensitivity required to
detect tidal features and the resolution required to resolve close pairs
($<5$\,kpc), making detailed morphological comparisons between ULIRGs
and SMGs impractical with current data.  For the 1-Jy ULIRG survey, 28\%
of ULIRGs show nuclear separations of more than 5\,kpc (Veilleux, Kim,
\& Sanders 2002).  In comparison, 2/7 (29\%) of the confirmed SMGs and
3/12 (25\%) of the confirmed plus candidate SMGs show double components
with similarly large separations.

The measured rest-frame optical properties of the SMGs are consistent
with those of low-redshift ULIRGs.  Previously the SMGs have been shown
to have similar infrared, CO, and radio properties to ULIRGs (e.g.,
Frayer et al. 1998, 1999, Ivison et al. 2000).  The NIR data suggest
that SMGs also have similar rest-frame optical properties.  The optical
luminosities, sizes, and morphologies of SMGs are consistent with
massive ($\ga M^{*}$) galaxies and major merger events and not
high-redshift sub-galactic clumps (e.g., Pascarelle et al. 1996).

\subsection{Extremely Red $J-K$ Galaxies}

Accounting for lensing (average amplification of 1\,mag), this survey
reached depths of approximately $K=23.5$ and $J=25.5$\,mag, which is as
deep as the deepest previous NIR surveys (Maihara et al. 2001; Franx et
al. 2003).  We observed 10 fields in both $J$- and $K$-band to
interesting depths, covering 4.4\,arcmin$^{2}$ on the sky.  Correcting
for lensing, the source-plane area surveyed is only about
1.8\,arcmin$^{2}$.  We detected 9 sources with $J-K>3$\,mag, yielding a
source density of 5\,arcmin$^{-2}$ at $K<22.5$\,mag ($J<25.5$\,mag).  In
comparison, the number density of $J-K>3$\,mag sources to similar depths
is about 1\,arcmin$^{-2}$ from the blank-field surveys (Totani et
al. 2001a, 2001b; Franx et al. 2003).  Given that SMGs tend to show
extremely red colors, it is not surprising to find a high number of
$J-K>3$\,mag sources in observations targeted at the positions of SMGs.
If we ignore all the $J-K>3$\,mag SMG candidate counterparts and
potential companions within 10$\arcsec$, the number density of extremely
red NIR sources decreases to about 2\,arcmin$^{-2}$ in the SMG fields.
Although roughly consistent with the blank-field result within the
errors associated with small number statistics, the slight excess may
indicate an enhanced density of ERO neighbors for the SMG population, as
recently suggested by Takata et al. (2003).

At a source density of 2\,arcmin$^{-2}$, the probability of randomly
finding a $J-K>3$\,mag source on the sky within the $15\arcsec$ SCUBA
beam is more than 10\%.  Therefore, the chance of finding an extremely
red source associated with a SCUBA detection is not negligible.  For the
strongest SCUBA detections ($\ga 5\sigma$), the 850$\mu$m centroid
positions are known to within about $\pm 3\arcsec$, which decreases the
probability of a chance positional coincidence to less than 1\%.  We
argue as a result that candidate sources with $J-K\ga3$\,mag within
5$\arcsec$ of the 850$\mu$m position indicate likely counterparts, but
that a red $J-K$ color by itself is not sufficient to confirm a SMG
counterpart.

The fraction of very and extremely red sources that are SMGs may
increase with redder colors.  Of the thirteen sources in the fields with
$J-K >2.5$\,mag, only five are candidate SMG counterparts, while four
out of the nine of the $J-K>3$\,mag sources are suspected SMGs (H5, J5,
N4, \& P7).  One of the $J-K>3$\,mag sources is arguably a companion to
H5 (B1, \S3.2).  The relationship of the remaining four $J-K>3$\,mag
sources to the SMG population is not clear.  As discussed by Totani et
al. (2001b), the density of SMGs is similar to that of extremely red NIR
sources, suggesting a possible connection between dusty SMGs and
galaxies with extremely red colors.  It is possible that the remaining
four $J-K>3$\,mag galaxies are simply SMGs below the 2-mJy 850$\mu$m
sensitivity limit of the SCUBA Cluster Lens Survey (corrected for
lensing).  At the fainter limit of 1\,mJy the number density of SMGs
(Blain et al.\ 2002) is sufficient to account for the number of
extremely red sources.

Based on this survey, SMGs with $J-K>3$\,mag account for $>40$\% of the
integrated 850$\mu$m background flux above 5\,mJy, correcting for
lensing.  Although many SMGs have extremely red $J-K$ colors, it is not
clear whether most $J-K>3$\,mag sources are infrared-bright SMGs.
Nearly all of the confirmed SMGs to date have extremely high
luminosities ($L\sim10^{13}\lsun$).  If the majority of the $J-K>3$\,mag
sources are examples of less luminous SMGs ($L\la 10^{12}\lsun$), then
their high counts would imply that this population may be responsible
for a large fraction of the star-formation activity at high redshift.
Alternatively, the majority of the faint $J-K>3$\,mag sources may have
much lower luminosities; it has even been proposed that many may be
Lyman break galaxies at $z\ga 10$ (e.g., Dickinson et al. 2000; Im et
al. 2002).  Observations with {\it SIRTF} will help to distinguish
between $z\ga10$ Lyman break systems and dust-reddened moderate redshift
systems from their spectral energy distributions.

\section{CONCLUSIONS}

Using deep NIR images of the SMGs in the SCUBA Cluster Lens Survey, we
find that a color selection criterion of $J-K\ga 3$\,mag can be used to
help identify candidate counterparts of SMGs that are too faint to be
detected at optical and radio wavelengths.  Faint sources ($K>20$\,mag)
with $J-K<2$\,mag can probably be ruled-out as SMG counterparts.

Of the 15 sources in the survey, all six of the brightest sub-mm sources
with intrinsic 850$\mu$m fluxes of $\ga$5\,mJy are identified at
1.4\,GHz and are moderately or very red in $J-K$.  The remaining 9
fainter sources lack strong radio detections, but we identify likely
counterparts to over half of these SMGs, based in part on their extreme
$J-K$ colors.  Only 3/15 of the sources in the survey remain
unidentified ($K>23$\,mag, corrected for lensing), all of which are
faint at sub-mm wavelengths.  These may be either more distant and/or
more obscured SMGs, or simply highly confused or spurious sources at
850$\mu$m.

The data on the SMG fields confirm the conclusion from blank-field
surveys that the fraction of extremely red NIR sources increases
dramatically at fainter magnitudes.  Within the SMG fields, we find a
number density of 5\,arcmin$^{-2}$ for sources with $J-K>3$\,mag
($K<22.5$\,mag), which is significantly higher than the number density
of bright SMGs.  It is not clear if the excess of extremely-red NIR
sources represents weaker infrared-luminous galaxies with
S(850$\mu$m)$\la 2$\,mJy, that are undetected by SCUBA, or if they
represent a different population of sources.

The median $K$-band magnitude for the SMG population is $K=22\pm1$\,mag,
and their median color is $J-K=2.6\pm0.6$\,mag.  Future observations
with {\it SIRTF} and ALMA will confirm or refute candidate counterparts.
The colors, rest-frame optical luminosities, sizes, and morphologies are
all consistent with the properties of low-redshift ULIRGs and the
association of SMGs with merger activity that may lead to the formation
of ellipticals or the bulges of galaxies.

\acknowledgements 

We thank the staff at the Keck Observatory who have made these
observations possible.  We are most fortunate to have the opportunity to
conduct observations from the summit of Mauna Kea which has a very
significant cultural role within the indigenous Hawaiian community.  We
acknowledge J. Surace, R. Ivison, J.-P. Knieb, L. Yan, and M. Im for
discussions about ULIRGs, SMGs, and EROs.  We thank D. Thompson for his
NIRCtools data reduction package and C. Frayer for proof reading the
manuscript.  D.T.F. and L.A.  are supported by the Jet Propulsion
Laboratory, California Institute of Technology, under contract with
NASA.  A.W.B., N.A.R., and I.R.S. acknowledge support from the NSF grant
AST-0205937, a NSF Graduate Research Fellowship, and both the Royal
Society and the Leverhulme Trust, respectively.

\begin{deluxetable}{lrrrr}
\tablenum{1}
\tablecaption{NIRC Observations}
\tablehead{\colhead{Source}&\colhead{S(850$\mu$m)}&\colhead{Submm}&\colhead{K-band}
&\colhead{J-band} \nl
&\colhead{(mJy)}&\colhead{$\alpha$,$\delta$(J2000)} &\colhead{(minutes)} &\colhead{(minutes)}}
              
\startdata

SMM\,J02399$-$0136  &23.0 & 02\,39\,51.9 $-$01\,35\,59 & 13  & 13 \nl
SMM\,J00266+1708    &18.6 & 00\,26\,34.1 $+$17\,08\,32 & 144 & 62 \nl
SMM\,J09429+4658    &17.2 & 09\,42\,54.7 $+$46\,58\,44 & 32  & 121\nl 
SMM\,J14009+0252    &14.5 & 14\,00\,57.7 $+$02\,52\,50 & 70  & 62 \nl
SMM\,J14011+0252    &12.3 & 14\,01\,05.0 $+$02\,52\,25 & 43  & 42 \nl
SMM\,J02399$-$0134  &11.0 & 02\,39\,56.4 $-$01\,34\,27 & 12  & 12 \nl
SMM\,J22471$-$0206  &9.2 & 22\,47\,10.2 $-$02\,05\,56 & 84  & 32 \nl
SMM\,J02400$-$0134  &7.6 & 02\,39\,57.8 $-$01\,34\,51 & 122 & 74 \nl
SMM\,J04431+0210    &7.2 & 04\,43\,07.2 $+$02\,10\,24 & 22  & 74 \nl
SMM\,J21536+1742    &6.7 & 21\,53\,38.4 $+$17\,42\,16 & 73  & 59 \nl
SMM\,J00265+1710    &6.1 & 00\,26\,31.3 $+$17\,10\,04 & 66  & 60 \nl
SMM\,J22472$-$0206  &6.1 & 22\,47\,13.9 $-$02\,06\,11 & 64  & 102\nl
SMM\,J00266+1710    &5.9 & 00\,26\,37.9 $+$17\,09\,51 & 104 & ...\nl
SMM\,J00267+1709    &5.0 & 00\,26\,39.7 $+$17\,09\,12 & 110 & ...\nl
SMM\,J04433+0210    &4.5 & 04\,43\,15.0 $+$02\,10\,02 & 100 & ...\nl

\enddata 

\tablenotetext{}{Notes--- The positions represent the location the
sub-mm emission, and the sources are listed in decreasing order of
850$\mu$m flux density (Smail et al. 2002).  Shorter NIRC integrations
were done for the bright sources (e.g., SMM\,J02399$-$0136 and
SMM\,J02399$-$0134), while longer observations were carried out for
fainter sources.}

\end{deluxetable}

\begin{deluxetable}{llccccrl}
\tablenum{2}
\tablecaption{Photometry Measurements}
\tablehead{\colhead{Field}&\colhead{Source} 
&\colhead{Offset}
&\colhead{Aperture}
&\colhead{K}
&\colhead{J}
&\colhead{J$-$K}
&\colhead{Comments}\nl 
&&\colhead{($\alpha\arcsec$,$\delta\arcsec$)}
&\colhead{Diameter}
&\colhead{(mag)} 
&\colhead{(mag)} 
&\colhead{(mag)} &}
              
\startdata
02399$-$0136 
&L1   & $-0.3$,1.0&3$\arcsec$ & 18.09$\pm$0.06 & 20.07$\pm$0.08 & 1.98
&\nl
&L2   & 1.9,0.5 &2$\arcsec$ & 20.29$\pm$0.13 & 22.43$\pm$0.15 & 2.14 & \nl
&L1+L2& ...    &6$\arcsec$ & 17.79$\pm$0.07 & 19.79$\pm$0.07 & 2.00 & SMG:CO\tablenotemark{a,b}\nl

00266+1708   
&M12  & 0.1,1.2& 2$\arcsec$ & 22.36$\pm$0.16 &  $>$24.27 & $>$1.91 &
SMG:mm\tablenotemark{c}\nl

09429+4658   
&H5 & $-0.5$,0.7 &2$\arcsec$ & 19.73$\pm$0.08 & $>$ 24.20 &$>$4.47 &
SMG:radio\tablenotemark{d}
\nl
&B1  & 2.7,3.3  &2$\arcsec$ & 20.48$\pm$0.12 & $>$ 24.32 &$>$3.84 & \nl

14009+0252   
&J5& $-$1.9,$-$0.9& 2$\arcsec$ & 20.52$\pm$0.09 & 23.87$\pm$0.21 & 3.35 & SMG:radio\tablenotemark{e}\nl

14011+0252   
&J1  &$-0.4$,$-0.4$ &6$\arcsec$ & 17.71$\pm$0.05 & 19.45$\pm$0.05 & 1.74 &
SMG:CO\tablenotemark{e,f,g,h}\nl
&J1a &$-0.4$,$-0.4$ &3$\arcsec$ & 17.86$\pm$0.06 & 19.62$\pm$0.06 & 1.76 & \nl
&J1n &0.2,1.7 &2$\arcsec$       & 19.89$\pm$0.13 & 21.76$\pm$0.11 & 1.87 & \nl
&J2  &$-2.4$,0.4&2$\arcsec$        & 19.76$\pm$0.07 & 21.39$\pm$0.07 & 1.63 & \nl
&J1+J2&...&6$\arcsec$       & 17.56$\pm$0.05 & 19.28$\pm$0.05 & 1.72 & \nl

02399$-$0134 
&L3& 1.6,$-0.1$ &6$\arcsec$ & 16.02$\pm$0.05 & 17.84$\pm$0.05 & 1.82 &
SMG:CO\tablenotemark{i,j}\nl

22471$-$0206 
&P4&$-1.5$,$-1.2$  &3$\arcsec$  &17.53$\pm$0.06 &  19.82$\pm$0.06 & 2.29 & SMG? \nl  
&P1a& 0.0,1.1  &2$\arcsec$ &20.41$\pm$0.08 & 22.01$\pm$0.08  & 1.60 & \nl
&P1b& $-1.2$,0.9 &2$\arcsec$ &21.02$\pm$0.10 & 23.01$\pm$0.12  & 1.99 & \nl
  
02400$-$0134 
&A1&$-3.0$,$-2.9$ &2$\arcsec$     &  20.73$\pm$0.08 &  23.59$\pm$0.17 & 2.86 & SMG? \nl  
&A2&1.7,$-0.9$ &2$\arcsec$     &  21.55$\pm$0.12 &  24.09$\pm$0.26 & 2.54 & \nl

04433+0210   
&N4 & 0.7,0.4& 2$\arcsec$ &19.41$\pm$0.09 & 22.56$\pm$0.28 & 3.15 &
SMG:CO\tablenotemark{k,l}\nl

21536+1742   
&K2&$-1.2$,$-3.0$  &2$\arcsec$ &  21.46$\pm$0.16 & 22.79$\pm$0.15  & 1.33 & \nl  
&K3   &$-0.8$,4.7 &6$\arcsec$ &  17.23$\pm$0.05 & 19.26$\pm$0.05  & 2.03 & SMG? \nl 
&K3a&$-0.6$,4.2 &3$\arcsec$ &  17.95$\pm$0.05 & 20.25$\pm$0.06  & 2.30 & \nl
&K3b&$-1.1$,5.5 &3$\arcsec$ &  18.41$\pm$0.05 & 20.21$\pm$0.06  & 1.80 & \nl

00265+1710   
& ...  &0.0,0.0&2$\arcsec$ &  $>$22.47 &  $>$22.85 &... & non-detection \nl
&M13 &3.5,3.3 &2$\arcsec$ & 19.41$\pm$0.07 &    20.96$\pm$0.07 &1.55 & \nl

22472$-$0206 
&P7&1.4,$-1.2$ &2$\arcsec$ & 21.05$\pm$0.13 &   $>$24.72 & $>$3.67 & SMG? \nl  
&P2&$-2.1$,3.2 &2$\arcsec$ & 21.93$\pm$0.25 &  23.78$\pm$0.17 & 1.85 & \nl
&A3&2.4,1.1 &2$\arcsec$ & 20.58$\pm$0.09 &  23.29$\pm$0.11  & 2.71 & \nl

00266+1710   
&...   &0.0,0.0&2$\arcsec$ &  $>$22.46 & ...  & ... & non-detection \nl
&M3 & $-3.2$,1.5  &2$\arcsec$ & 21.10$\pm$0.11 & ... & ... & \nl

00267+1709   
&...  &0.0,0.0&2$\arcsec$ &  $>$21.85 & ...  & ... & non-detection \nl

04433+0210   
&N5  & 0.9,$-0.8$& 2$\arcsec$ & 22.43$\pm$0.24 & ...  & ... & SMG:radio?  \nl

\enddata 

\tablenotetext{}{Notes--- Offsets are with respect to the sub-mm
 positions given in Table~1.  Lower limits are $3\sigma$.  The candidate
 SMGs are shown with a ``?''.  The confirmed SMGs are listed without
 question marks and are discussed in the following references: $^{\rm
 a}$Ivison et al.\ (1998), $^{\rm b}$Frayer et al.\ (1998), $^{\rm
 c}$Frayer et al.\ (2000), $^{\rm d}$Smail et al.\ (1999), $^{\rm
 e}$Ivison et al.\ (2000), $^{\rm f}$Frayer et al.\ (1999), $^{\rm
 g}$Ivison et al.\ (2001), $^{\rm h}$Downes \& Solomon(2003), $^{\rm
 i}$Soucail et al.\ (1999), $^{\rm j}$J.-P. Kneib, private
 communication, $^{\rm k}$Frayer et al.\ (2003), $^{\rm l}$R. Neri et
 al.\ (2003).}

\end{deluxetable}

\newpage
\begin{figure}[t]
\includegraphics{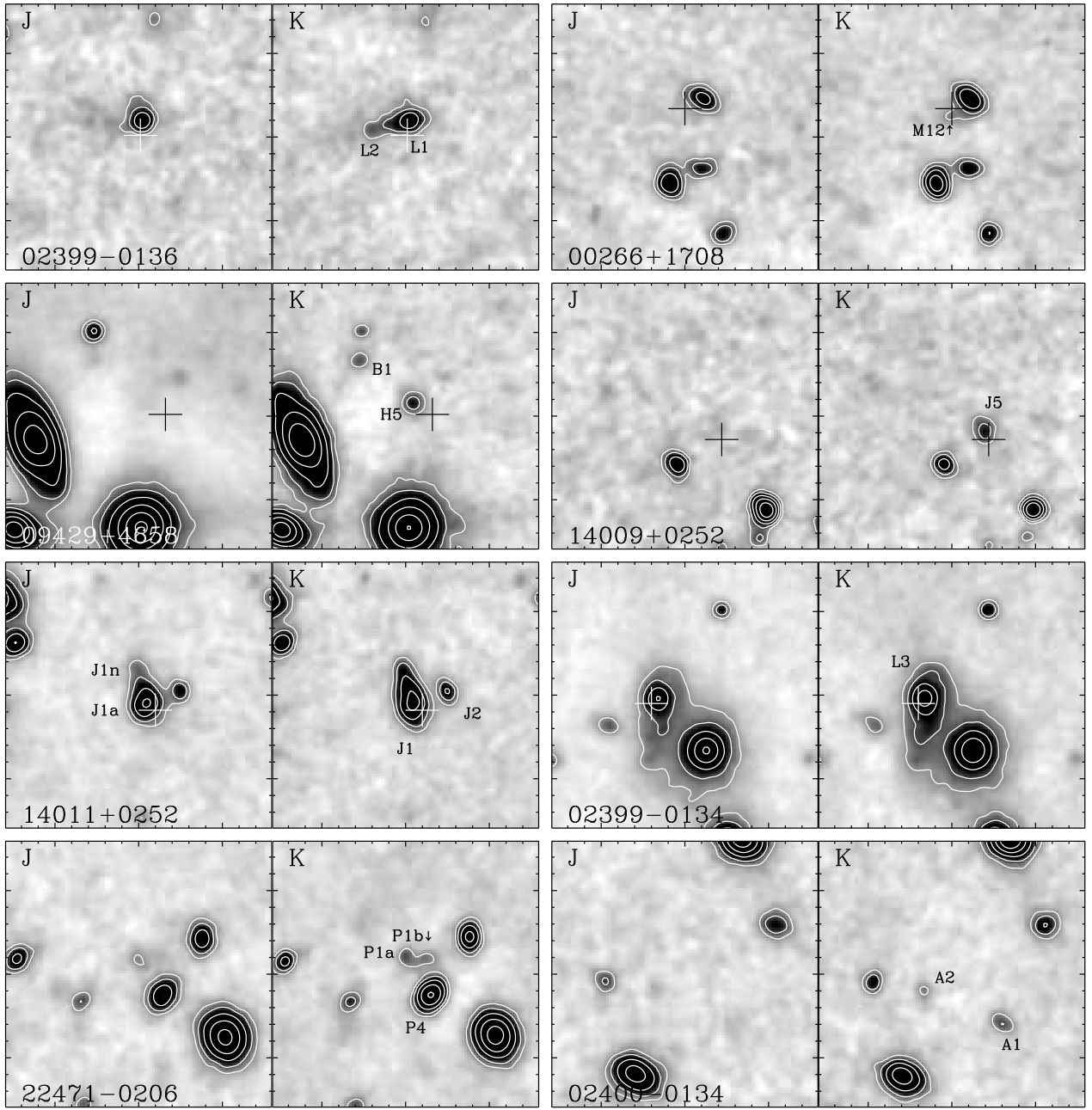}
\vspace*{7.0in}
\caption{NIRC $J$-band ($1.2\mu$m) and $K$-band
($2.2\mu$m) images of the SMG fields (North is up and East is left).
The images are $16\arcsec \times 16 \arcsec$ with tick marks every
$1\arcsec$.  The images are centered on the position of the 850$\mu$m
emission given in Table~1. For the confirmed SMG counterparts, crosses
indicate the position of the CO, mm, or radio emission.  The data have
been smoothed with a $0\farcs5$ FWHM Gaussian.  The grey-scale is plotted
on a logarithmic scale, from $-3\sigma$ (white) to $+15\sigma$ (black),
and the contours start at $+5\sigma$ and increase by factors of 2.}
\end{figure}

\newpage
\setcounter{figure}{0}

\begin{figure}[t]
\includegraphics{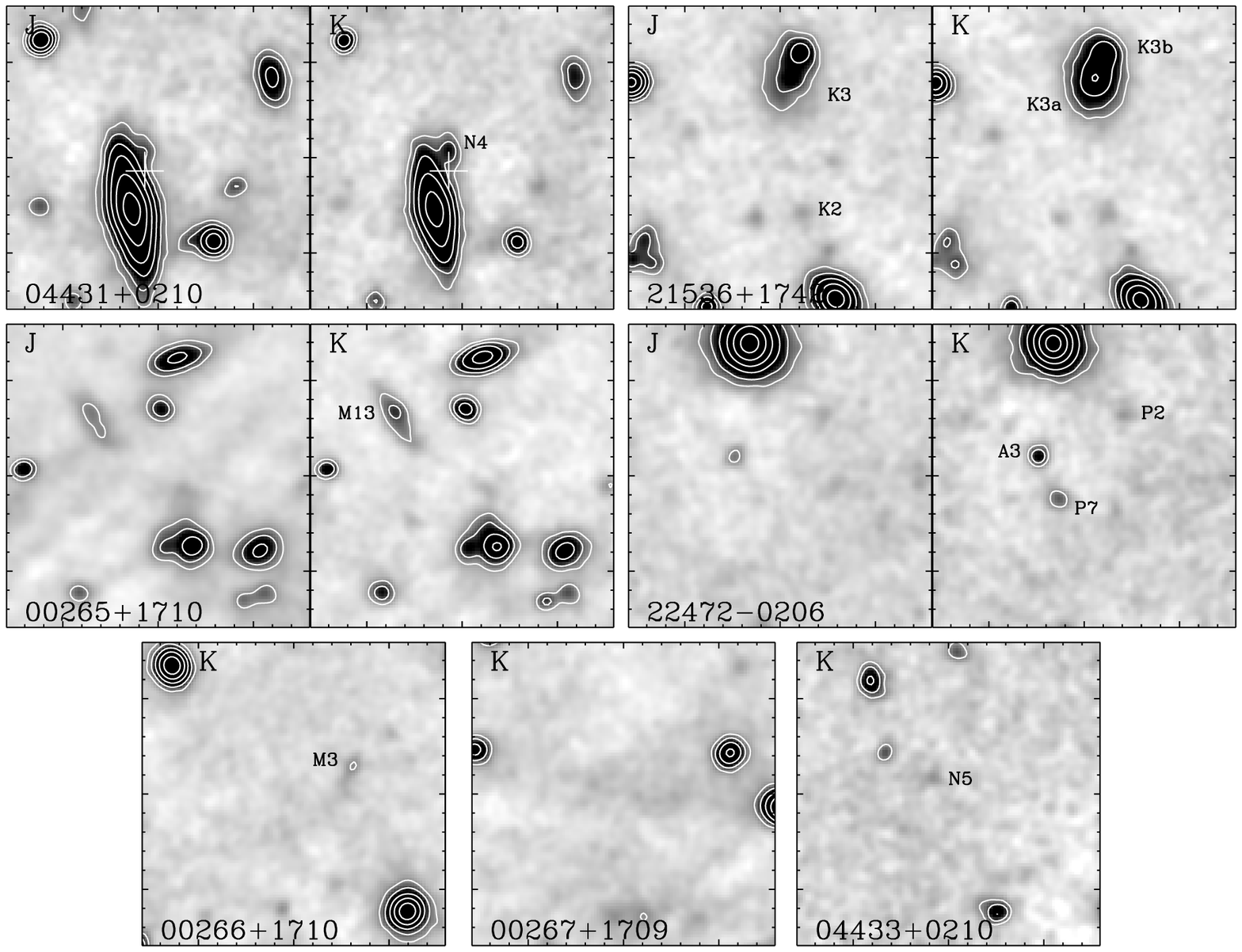}
\vspace*{5.0in}
\caption{Continued.  The last three fields only have
$K$-band data.}
\end{figure}

\newpage
\begin{figure}[t]
\includegraphics{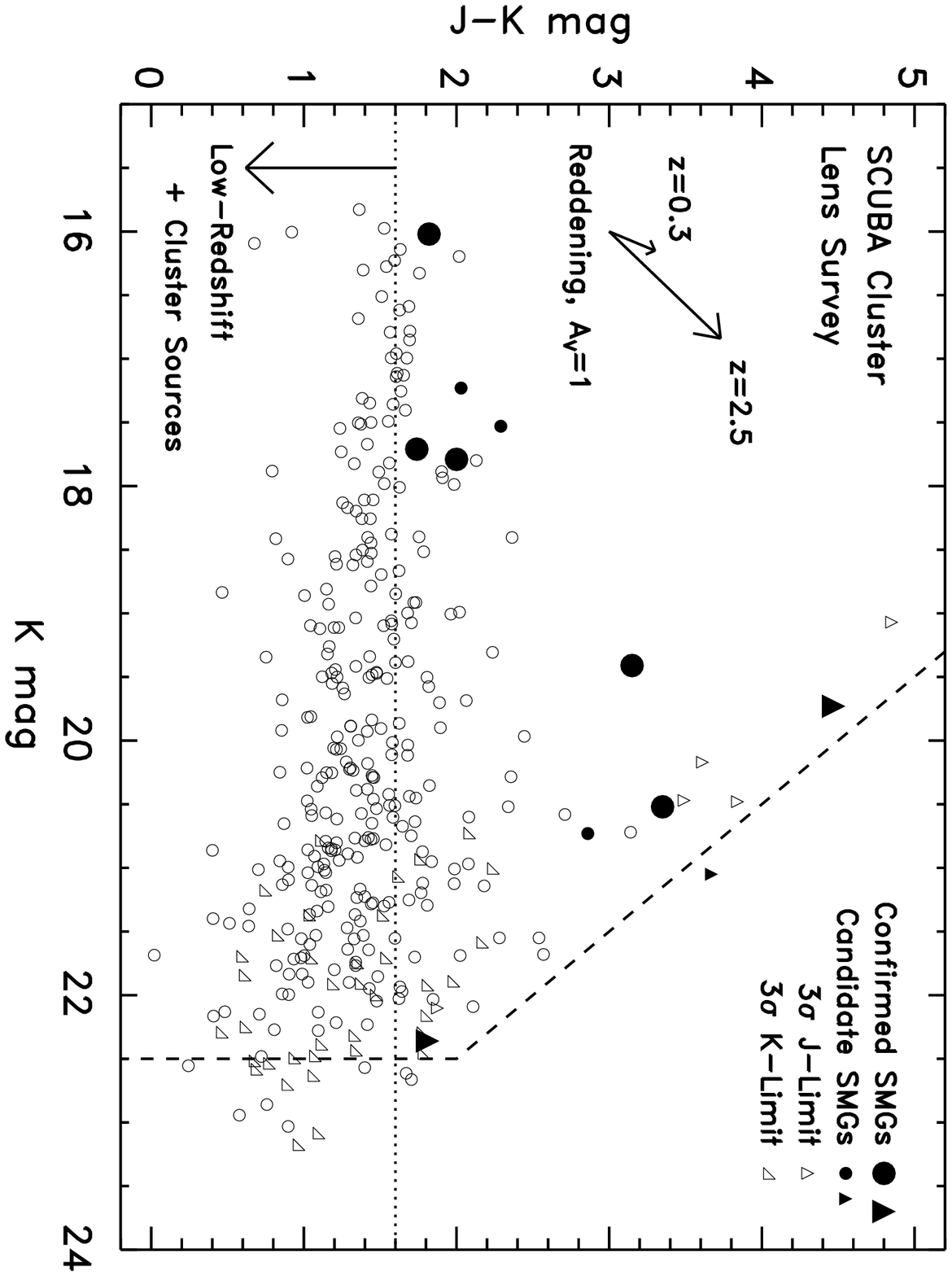}
\vspace*{5.0in}
\caption{Photometry results for all 339 detected
sources in the 12 SMG fields with $J$- and $K$-band observations.  The
confirmed and candidate SMG counterparts are plotted as large and small
solid symbols, respectively.  Circles represent detections in both $J$-
and $K$-band, upward pointing triangles show $K$-band detections with
$3\sigma$ $J$-band limits, and triangles pointed toward the lower-right
show $J$-band detections with $3\sigma$ $K$-band limits.  Almost all
sources below the dotted line ($J-K<1.6$\,mag) are expected to be
low-redshift sources and cluster galaxies.  The dashed line represents
the survey limits of $K<22.5$\,mag and $J<24.5$\,mag ($3\sigma$).  The
background SMGs would need to be shifted about one magnitude fainter in
$K$-band on average to correct for lensing.  Reddening vectors are shown
for a source-frame visual extinction of A$_V=1$\,mag at the typical
cluster ($z=0.3$) and SMG ($z=2.5$) redshifts, assuming the standard
Galactic interstellar extinction law (Whitford 1958).}
\end{figure}


\begin{references}

\reference{} Barger, A. J., Cowie, L. L., Richards, E. A. 2000, AJ, 119,
2092

\reference{} Barger, A. J., Cowie, L. L., \& Sanders, D. B.\ 1999a, ApJ,
518, L5

\reference{} Barger, A. J., Cowie, L. L., Smail, I., Ivison, R. J.,
Blain, A. W., \& Kneib, J.-P. 1999b, AJ, 117, 2656

\reference{} Bertin E. \& Arnouts S., 1996, A\&AS, 117, 393 

\reference{} Blain, A. W., Barnard, V. E., \& Chapman, S. C. 2003,
MNRAS, 338, 733

\reference{} Blain, A. W., Smail, I., Ivison, R. J., Kneib, J.-P., \&
Frayer, D. T.\ 2002, Physics Reports, 369, 111

\reference{} Chapman, S. C. et al. 2003a, ApJ, 585, 57

\reference{} Chapman, S. C., Blain, A. W., Ivison, R. J., Smail,
I. R. 2003b, Nature, 422, 695

\reference{} Chapman, S. C., Richards, E. A., Lewis, G. F., Wilson, G.,
Barger, A. J. 2001, ApJ, 548, L147

\reference{} Chapman, S. C., Scott, D., Borys, C., \& Fahlman,
G. G. 2002a, MNRAS, 330, 92

\reference{} Chapman, S. C., Smail, I., Ivison, R. J., Helou, G., Dale, D. A.,
\& Lagache, G. 2002b, ApJ, 573, 66

\reference{} Cowie, L. L., Barger, A. J., Kneib, J.-P, 2002, AJ, 123,
2197

\reference{} Dickinson, M. et al. 2000, ApJ, 531, 624

\reference{} Downes, D., \& Solomon, P. M. 2003, ApJ, 582, 37

\reference{} Eales, S., Lilly, S., Gear, W., Dunne, L., Bond, J. R.,
Hammer, F., Le F\`{e}vre, O., \& Crampton, D.\ 1999, ApJ, 515, 518

\reference{} Eales, S., Lilly, S., Webb, T., Dunne, L., Gear, W.,
 Clements, D., Yun, M.\ 2000, AJ, 120, 2244

\reference{} Franx, M., et al. 2003, ApJ, 587, L79

\reference{} Frayer, D. T., Armus, L., Scoville, N. Z., Blain, A. W.,
Reddy, N. A., Ivison, R. J., \& Smail, I. 2003, AJ, 126, 73
 
\reference{} Frayer, D. T., et al.\ 1999, ApJ, 514, L13

\reference{} Frayer, D. T., Ivison, R., J., Scoville, N. Z., Yun, M.,
Evans, A. S., Smail, I., Blain, A. W., \&, Kneib, J.-P.\ 1998, ApJ, 506,
L7

\reference{} Frayer, D. T., Smail, I., Ivison, R., J., \& Scoville,
N. Z.\ 2000, AJ, 120, 1668

\reference{} Hughes, D., et al.\ 1998, Nature, 394, 241

\reference{} Im, M., Yamada, T., Tanaka, I., \& Kajisawa, M. 2002, ApJ,
578, L19
  
\reference{} Ivison, R. J., Smail, I., Barger, A. J., Kneib, J.-P.,
Blain, A. W., Owen, F. N., Kerr, T. H., \& Cowie, L. L.\ 2000, MNRAS,
315, 209

\reference{} Ivison, R. J., et al. 2002, MNRAS, 337, 1

\reference{} Ivison, R. J., Smail, I., Frayer, D. T., Kneib, J.-P., \&
Blain, A. W.\ 2001, ApJ, 561, L45

\reference{} Ivison, R. J., Smail, I., Le Borgne, J.-F., Blain, A. W.,
Kneib, J.-P., B\'{e}zecourt, J., Kerr, T. H., \& Davies, J. K.\ 1998,
MNRAS, 298, 583

\reference{} Kim, D.-C., Veilleux, S., \& Sanders, D. B.\ 2002, ApJ,
143, 277

\reference{} Lilly, S. J., Eales, S. A., Gear, W. K. P., Hammer, F., Le
F\`{e}vre, O., Crampton, D., Bond, J. R., \& Dunne, L. 1999, ApJ, 518, 641

\reference{} Maihara, T., et al. 2001, PASJ, 53, 25

\reference{} Matthews, K., \& Soifer, B. T.\ 1994, Infrared Astronomy
with Arrays: the Next Generation, ed. I. McLean (Dordrecht: Kluwer
Academic Publishers), 239

\reference{} Metcalfe, L. et al. 2003, A\&A, 407, 791

\reference{} Murphy, T. W., Jr., Armus, L., Matthews, K., Soifer, B. T.,
Mazzarella, J. M., Shupe, D. L., Strauss, M. A., Neugebauer, G. 1996,
AJ, 111, 1025

\reference{} Neri, R., et al.\ 2003, ApJL, in press (astro-ph/0307310)

\reference{} Pascarelle, S. M., Windhorst, R. A., Keel, W. C., \&
Odewahn, S. C. 1996, Nature, 383, 45

\reference{} Persson, S. E., Murphy, D. C., Krzeminski, W., Roth, M., \&
Rieke, M. J.\ 1998, AJ, 116, 2475

\reference{} Scott, S. E., et al.\ 2002, MNRAS, 331, 817

\reference{} Scoville, N. Z., et al. 2000, AJ, 119, 991 

\reference{} Shapley, A. E., Steidel, C. C., Adelberger, K.  L.,
Dickinson, M., Giavalisco, M., Pettini, M. 2001, ApJ, 562, 95

\reference{} Smail, I., Ivison, R. J., \& Blain, A. W.\ 1997, ApJ, 490, L5

\reference{} Smail, I., Ivison, R. J., Blain, A. W., \& Kneib, J.-P.\
2002, MNRAS, 331, 495

\reference{} Smail, I., Ivison, R. J., Blain, A. W., \& Kneib, J.-P.\
1998, ApJ, 507, L21

\reference{} Smail, I., Ivison, R. J., Kneib, J.-P., Cowie, L. L.,
Blain, A. W., Barger, A. J., Owen, F. N., \& Morrison, G.\ 1999, MNRAS,
308, 1061

\reference{} Smail, I., Ivison, R. J., Owen, F. N., Blain, A. W., \&
Kneib, J.-P.\ 2000, ApJ, 528, 612

\reference{} Soucail, G., Kneib, J. P., B\'{e}zecourt, J., Metcalfe, L.,
Altieri, B., Le Borgne, J. F.\ 1999, A\&A, 342, L70

\reference{} Surace, J. A., \& Sanders, D. B.\ 2000, AJ, 120, 604

\reference{} Surace, J. A., Sanders, D. B., Vacca, W. D.,
Veilleux, S., Mazzarella, J. M. 1998, ApJ, 492, 116

\reference{} Takata, T., et al. 2003, PASJ, 55, 789

\reference{} Totani, T., Yoshii, Y., Iwamuro, F., Maihara, T., \&
Motohara, K. 2001a, ApJ, 559, 592

\reference{} Totani, T., Yoshii, Y.,  Maihara, T., Iwamuro, F., \&
Motohara, K. 2001b, ApJ, 558, L87

\reference{} Veilleux, S., Kim, D.-C., \& Sanders, D. B.\ 2002, ApJ,
143, 315

\reference{} Webb, T. M., et al.\ 2003, ApJ, 587, 41

\reference{} Wehner, E. H., Barger, A. J., \& Kneib, J.-P. 2002, ApJ,
577, L83

\reference{} Whitford, A. E. 1958, AJ, 63, 201

\end{references}
\end{document}